\documentclass[aps,pre,twocolumn,superscriptaddress,showpacs]{revtex4-1}
\pdfoutput=1
\usepackage{graphicx}
\usepackage{amsmath}
\begin{document}


\title{A neuronal network model of interictal and recurrent ictal activity}


\author{M. A. Lopes}
\email{m.lopes@exeter.ac.uk}
\affiliation{College of Engineering, Mathematics and Physical Sciences, University of Exeter, Devon EX4, United Kingdom}
\affiliation{Wellcome Trust Centre for Biomedical Modelling and Analysis, University of Exeter, Devon EX4, United Kingdom}
\affiliation{EPSRC Centre for Predictive Modelling in Healthcare, University of Exeter, Devon EX4, United Kingdom}
\affiliation{Department of Physics $\&$ I3N, University of Aveiro, 3810-193 Aveiro, Portugal}
\author{K.-E. Lee}
\affiliation{Department of Physics $\&$ I3N, University of Aveiro, 3810-193 Aveiro, Portugal}
\affiliation{Department of Anesthesiology and Center for Consciousness Science, University of Michigan Medical School, Ann Arbor, Michigan, USA}
\author{A. V. Goltsev}
\affiliation{Department of Physics $\&$ I3N, University of Aveiro, 3810-193 Aveiro, Portugal}
\affiliation{A.F. Ioffe Physico-Technical Institue, 194021 St. Petersburg, Russia}




\begin{abstract}
We propose a neuronal network model which undergoes a saddle-node on an invariant circle bifurcation as the mechanism of the transition from the interictal to the ictal (seizure) state. In the vicinity of this transition, the model captures important dynamical features of both interictal and ictal states. We study the nature of interictal spikes and early warnings of the transition predicted by this model. We further demonstrate that recurrent seizures emerge due to the interaction between two networks.
\end{abstract}

\pacs{87.19.lj,87.19.xm,05.70.Fh,05.10.-a,87.19.ln,87.18.Sn}   
\maketitle

\section{Introduction}
\label{introduction}
Epilepsy affects nearly $1 \%$ of the population worldwide \cite{Banerjee_2009}. This disabling neurological disorder is characterized by spontaneous recurrent seizures, which correlate to strongly synchronized neuronal activities, so-called paroxysmal activity, revealed in electroencephalograms (EEG). Anticonvulsant medications can prevent seizures, but side effects are frequently reported \cite{Ortinski_2004}. For about $30 \%$ of the patients, medications are not effective \cite{Kwan_2000}. A minority undergoes surgery to remove the epileptogenic brain tissue, but even in these cases the patients may continue experiencing spontaneous seizures \cite{Tisi_2011,Najm_2013}. One of the main challenges has been to try to forecast seizures \cite{Chavez_2003,Quyen_2005,Mormann_2007}. On one hand, the unpredictability of seizure occurrence is a major burden of the condition \cite{Boer_2008} and therefore being able to alert patients of impending seizures could greatly improve their quality of life. On the other hand, it would allow the design of closed-loop intervention systems which could stop seizures \cite{Stacey_2008}. Much of the research in seizure prediction has been focused on algorithms \cite{Mormann_2007}, however a better understanding of epilepsy mechanisms is required.

Although epilepsy is an umbrella term for a range of syndromes, the electrophysiological signatures are similar between them \cite{Jirsa_2014}. For instance, different epileptogenic lesions can produce similar electroencephalographic patterns \cite{Perucca_2014}. Also, it is noteworthy that it is possible to induce seizures in non-epileptic brains across species both \textit{in vivo} and \textit{in vitro}, which again present similar electrophysiological features (see e.g. \cite{Raol_2011,Huberfeld_2011}). As argued by Jirsa \textit{et al.} \cite{Jirsa_2014}, these facts suggest the existence of invariant dynamical properties underlying seizure dynamics. Moreover, there are evidences that seizures self-terminate via a critical transition \cite{Kramer_2012}. Note that bifurcations are the mechanisms of phase transitions in many-body interacting systems \cite{Strogatz_1994,Scheffer_2009}. The two frameworks provide complementary insights about the underlying transitions. While there are extensive studies on what kind of bifurcations occur at the onset and offset of seizures (e.g. \cite{Jirsa_2014,Breakspear_2006,Suffczynski_2004}), there are fewer studies about the collective nature of the phase transitions \cite{Steyn-Ross_2010}. The fundamental question is how does the spike interaction between large populations of neurons may result in seizures.

Critical phenomena provide early-warnings of phase transitions \cite{Scheffer_2009}, which consequently open the possibility to take action to prevent the occurrence of those transitions. We have previously demonstrated that the interaction between neurons on a network give rise to collective phenomena and diverse phase transitions \cite{Lee_2014}. Note that different phase transitions are associated with different precursors and different critical phenomena such as (among other) bursts of neuronal activity, avalanches, hysteresis, critical slowing-down, symmetry breaking, and resonance phenomena \cite{Lee_2014,Lopes_2014}. 
 
Here, we propose a neuronal network model consisting  of interacting excitatory and inhibitory neurons to understand the nature and emergence of both interictal and ictal activity. We start by presenting the model and its dynamical states. We then study the properties of interictal-like spikes, which are evoked in the vicinity of the transition to the ictal state. As the dynamical state moves towards the transition, early-warning phenomena signal the impending transition. We demonstrate that such phenomena is revealed through the stimulation of interictal-like spikes, and the analysis of accompanying low-fluctuating activity. Finally, we show that recurrent ictal activity is an emergent collective phenomenon in a system of two interacting neuronal networks.

\section{Neuronal network model}
Herein we consider a neuronal network model \cite{Goltsev_2010,Lee_2014,Lopes_2014}, which we will refer to as the stochastic cellular automata neuronal network model (SCANNM). In the SCANNM, neurons are modelled as stochastic integrate-and-fire neurons: they integrate the inputs and fire a train of spikes with a certain probability if the input is larger than an activation threshold (see Appendix \ref{appendix_SCANNM} for more details). There are two populations of neurons, excitatory and inhibitory neurons. Excitatory neurons fire positive outputs, whereas inhibitory fire negative outputs to their postsynaptic neighbours. In general, the two populations have different response times to stimuli. Here we consider the case in which excitatory neurons respond faster than inhibitory neurons. Additionally, neurons are excited by endogenous stimuli that account for random spikes coming from other areas of the brain, as well as spontaneous releases of neurotransmitters at the synapses. $\langle n \rangle$ is the endogenous stimulation and we use it as control parameter. The neurons form an uncorrelated random directed network (properties of this kind of networks have been studied, for example, in \cite{Dorogovtsev_2002,Newman_2003,Boccaletti_2006}). The mean-field neuronal dynamics are described by the fractions of active excitatory ($\rho_e$) and inhibitory ($\rho_i$) neurons which follow the rate equations \cite{Goltsev_2010,Lee_2014}
\begin{equation}
\frac{\dot{\rho_a}}{\mu_a}=-\rho_a+\Psi_a(\rho_e,\rho_i,\langle n \rangle),
\label{rho-eq}
\end{equation}
where $a=e,i$, and $\dot{\rho} \equiv d\rho /dt$. $\Psi_a(\rho_e,\rho_i,\langle n \rangle)$ is the probability that at time $t$ a randomly chosen excitatory ($a=e$) or inhibitory ($a=i$) neuron becomes active. This function encodes the network structure, single neuron stochastic firing rules, and endogenous stimulation \cite{Goltsev_2010,Lee_2014,Lopes_2014}.

The model is analytically solvable, but despite its simplicity, it describes a rich repertoire of collective phenomena, namely neuronal avalanches, bursty activity, hysteresis, bistability, different kinds of neuronal oscillations, phase transitions, and stochastic resonance \cite{Lee_2014,Lopes_2014}. Furthermore, the SCANNM combines two usually distinct modelling frameworks to describe mesoscopic brain dynamics, namely it allows the modelling of large-scale neuronal networks like in Ref.~\cite{Izhikevich_2008}, and it is simultaneously described by a neural mass formulation (see for instance \cite{Stefanescu_2012}). It thus enables an analysis of both single neuron dynamics within the network, and large-scale dynamics of neuronal populations that can be treated numerically and analytically. 

\begin{figure}
\includegraphics[width=0.45\textwidth]{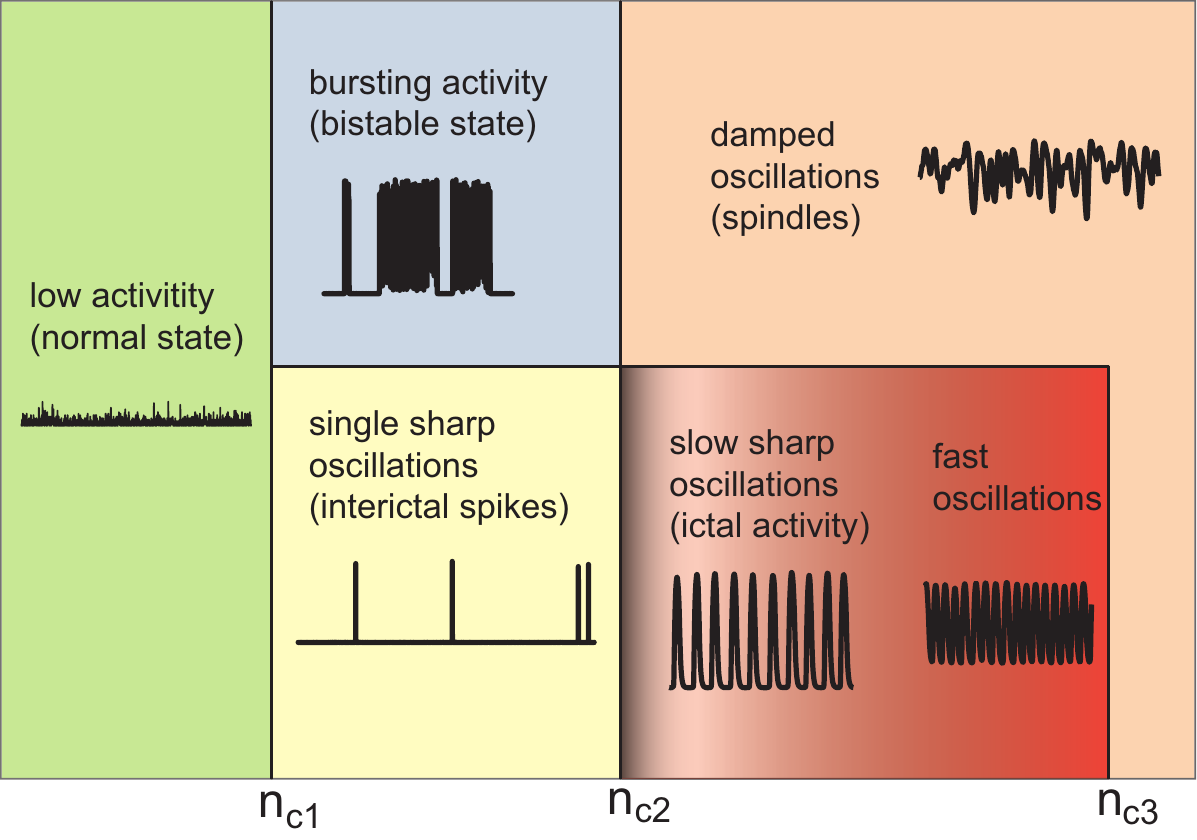}
\caption{Schematic diagram of the different dynamical states of the SCANNM as function of the endogenous stimulation. 
We find low neuronal activity at low stimulation ($\langle n \rangle < n_{c1}$). At an intermediate stimulation level ($n_{c1} < \langle n \rangle < n_{c2}$), the neuronal network exhibits transients of high activity, either bursts or single sharp oscillations, which give place to damped oscillations or sustained oscillations at $\langle n \rangle > n_{c2}$ respectively depending on parameters. The shape of the sustained oscillations changes from slow, high-amplitude oscillations to fast, low-amplitude oscillations as the stimulation increases. The neuronal network produces damped oscillations to a high activity state at $\langle n \rangle > n_{c3}$. 
\label{diagram-fig}}
\end{figure}

Figure~\ref{diagram-fig} depicts the different patterns of neuronal activity in the SCANNM. In this paper we focus on the regions corresponding to (i) low fluctuating activity around a stable state, which we identify as a 'normal' state; (ii) low fluctuating activity with sporadic single sharp oscillations, the 'interictal' state; and (iii) sustained network oscillations, which is the model 'ictal' state. The boundary between the interictal and ictal regions corresponds to a saddle-node on an invariant circle (SNIC) bifurcation (at $\langle n \rangle=n_{c2}$), which is the critical point of a second-order phase transition from the interictal to the ictal state \cite{Lee_2014}. At higher endogenous stimulations, the ictal region is bounded by the critical point of a supercritical Hopf bifurcation that separates the ictal state from a high activity state (at $\langle n \rangle=n_{c3}$). In the vicinity of the Hopf bifurcation, the neuronal oscillations have high frequency, and low amplitude, in contrast to the oscillations close to the SNIC bifurcation, which are characterized by low-frequency and high-amplitude \cite{Izhikevich_2000}. In the interictal state, interictal-like spikes (ILS) emerge at random, but with a deterministic shape. Note that ILS are strongly nonlinear events that comprise the synchronous activity of almost $90 \%$ of the neurons in the network. An ILS is described by a trajectory that goes around an unstable point in the $(\rho_e,\rho_i)-$phase plane (see Sec.~\ref{properties}). Their occurrence is deterministic if the activity overcomes a threshold (a separatrix in the phase plane). This threshold defines the number of excitatory neurons that must be activated simultaneously in order to generate an ILS. As an example, for a network of $10^4$ neurons at $\langle n \rangle= 16$ (i.e., at $(n_{c2}-\langle n \rangle) / n_{c2} \approx 0.15$), which is in a low activity state where almost all neurons are inactive, the simultaneous activation of just $75$ excitatory neurons chosen at random (i.e., about $1 \%$ of the excitatory neurons in a network with $25 \%$ inhibitory neurons) generates an ILS formed by the synchronized activity of about $9000$ neurons \cite{Lee_2014}. One should note that the duration of these ILS is much larger than the period of single neuron spikes. This implies that an ILS is genuinely a collective phenomena in spite of the fact that it can be elicited by a small number of neurons. For realistic parameters, namely mean degree $c=1000$, fraction of inhibitory neurons $g_i=0.25$, synaptic efficacies ratio $J_{i} / J_{e}=-3$, dimensionless firing threshold $\Omega=30$, ratio between excitatory and inhibitory response times $\alpha=\mu_i / \mu_e = 0.7$, and $1 / \mu_e  = 20$ ms (see Appendix \ref{appendix_SCANNM} and Refs. \cite{Goltsev_2010,Lee_2014,Lopes_2014}), the typical duration of an ILS is about $100$ ms which is comparable to real interictal spikes \cite{Karoly_2016}. Within the ictal region, the frequency of sustained oscillations increases with endogenous stimulation from very low frequencies up to $4$ Hz \cite{Lee_2014}, which is comparable to the frequency of ictal activity \cite{Markand_2003}. These are natural features of the model, without needing to calibrate parameters.

\section{The nature of the interictal-like spikes} 
\label{properties}
\begin{figure}
\includegraphics[width=0.3\textwidth]{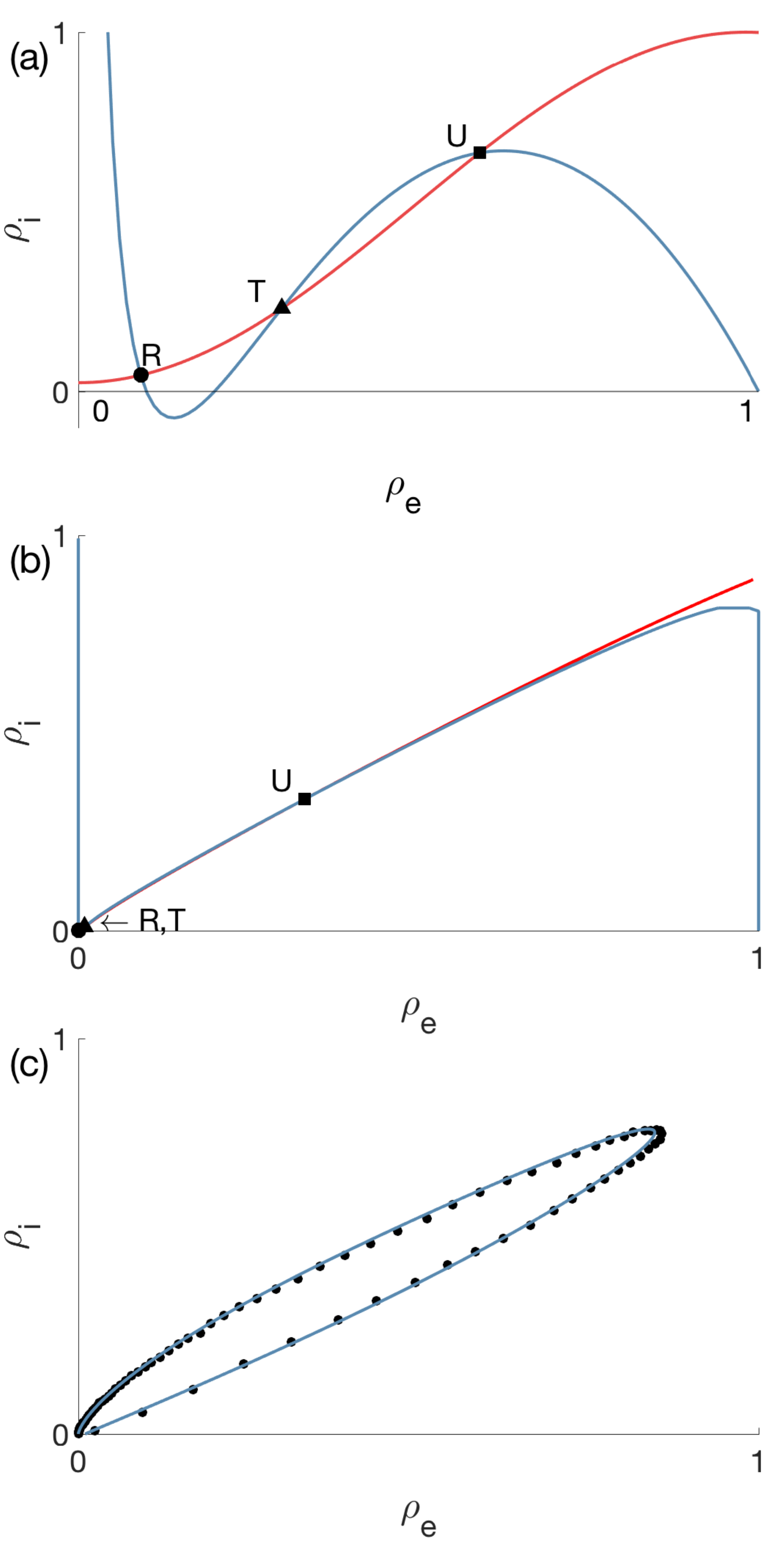}
\caption{(a) Schematic representation of the nullclines and fixed points in the $(\rho_e,\rho_i)-$phase plane. 
(b) Nullclines and fixed points of the SCANNM (numerical integration of Eqs.~(\ref{null-eq})).
In panels (a) and (b) the blue and red lines correspond to the $\rho_e$- and $\rho_i$-nullcline, respectively. The nullclines intersect at the fixed points: the stable or rest state (R), the saddle or threshold (T), and the unstable point (U). 
(c) Trajectory of an ILS in the $(\rho_e,\rho_i)-$phase plane. The line is the result of the numerical integration of Eqs.~(\ref{rho-eq}) and the dots represent simulations of the model. 
Parameters: $c=1000$, $\Omega=30$, $g_i=0.25$, $J_i=-3J_e$, $\sigma^2=10$, $\langle n \rangle=16$, $\alpha=0.7$, $\tau=0.1$ and $N=10^4$ (see the Appendix \ref{appendix_SCANNM} for details about the parameters). 
\label{spike_nullclines-fig}}
\end{figure}

In order to understand the nature of ILS, let us study their phase trajectories in the plane $\rho_e-\rho_i$. ILS emerge in the region $n_{c1} < \langle n \rangle < n_{c2}$ (see Fig.~\ref{diagram-fig}). In this region there are three fixed points: a stable fixed point at low activity, a saddle point at an intermediate activity, and an unstable point at high activity \cite{Lee_2014}. For this parameter region, the dynamics in the SCANNM is qualitatively equivalent to the dynamics of the Morris-Lecar neuron near the SNIC bifurcation \cite{Rinzel_1989}. To the best of our knowledge, a SNIC bifurcation has not been found in any other neuronal network model. Thus, the SCANNM gives an unique possibility to study the collective behavior of neuronal populations near a SNIC bifurcation. We follow the nomenclature used by Rinzel and Ermentrout \cite{Rinzel_1989}: the stable point corresponds to a rest state (R), and the saddle point is a threshold (T) on the separatrix that divides the phase plane into two regions. There are two heteroclinic orbits connecting T to R, one corresponding to immediate exponential relaxation to R, and another that goes around the unstable point (U), reaching high activity before relaxing to R. The ILS follows the second path (see Fig.~\ref{spike_nullclines-fig}) as spikes of single neurons do in the Morris-Lecar model. Panels (a) and (b) of Fig.~\ref{spike_nullclines-fig} also portray the nullclines of the system,
\begin{equation}
\dot{\rho_a}=0 \Leftrightarrow \rho_a=\Psi_a(\rho_e,\rho_i,\langle n \rangle).
\label{null-eq}
\end{equation} 
The nullclines determine the maximums or minimums of excitatory and inhibitory activity. In this case, the activity of the population $a$ increases ($\dot{\rho}_a>0$) below the respective nullcline, whereas it decreases above. Consequently, the ILS move counterclockwise in the phase plane. Thus, any excitatory activity perturbation that drives the activity state below both nullclines results in an ILS as the one displayed in Fig.~\ref{spike_nullclines-fig}(c). At the critical point $n_{c2}$, the points R and T merge, and there is a homoclinic orbit around the unstable point connecting the saddle-node to itself. 

\begin{figure}
\includegraphics[width=0.45\textwidth]{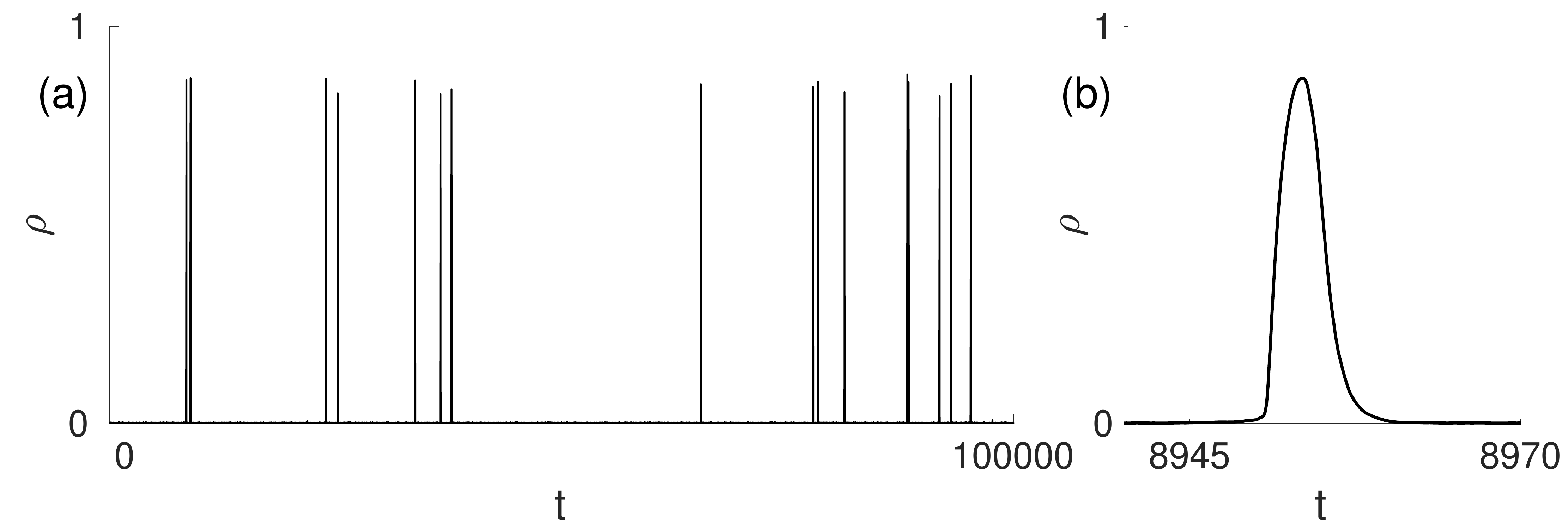}
\caption{(a) Series of ILS generated at random due to finite-size effects at $\langle n \rangle=18.75$ ($n_{c2}=18.8$). 
(b) Zoom of an ILS from panel (a). All ILS have the same shape. 
Parameters in simulations are the same as in Fig.~\ref{spike_nullclines-fig}.
\label{spike_series}}
\end{figure}

The distance between R and T defines an activation threshold $A_{th}(\langle n \rangle)$ for the generation of ILS. We demonstrate in Appendix \ref{app_activation_th} that $A_{th}(\langle n \rangle)$ follows a square root dependence with $\langle n \rangle$ in the vicinity of the SNIC bifurcation,
\begin{equation}
A_{th}(\langle n \rangle) \propto \sqrt{n_{c2}-\langle n \rangle}.
\label{activation-eq}
\end{equation}
In a finite network, finite-size effects elicit activity fluctuations which can overcome the activation threshold provided that the system is sufficiently close to the critical point, which in turn results in the occasional generation of ILS (see Fig.~\ref{spike_series}). 

\section{Early warnings of the transition to the ictal state}
\label{forecast}
\begin{figure}
\includegraphics[width=0.45\textwidth]{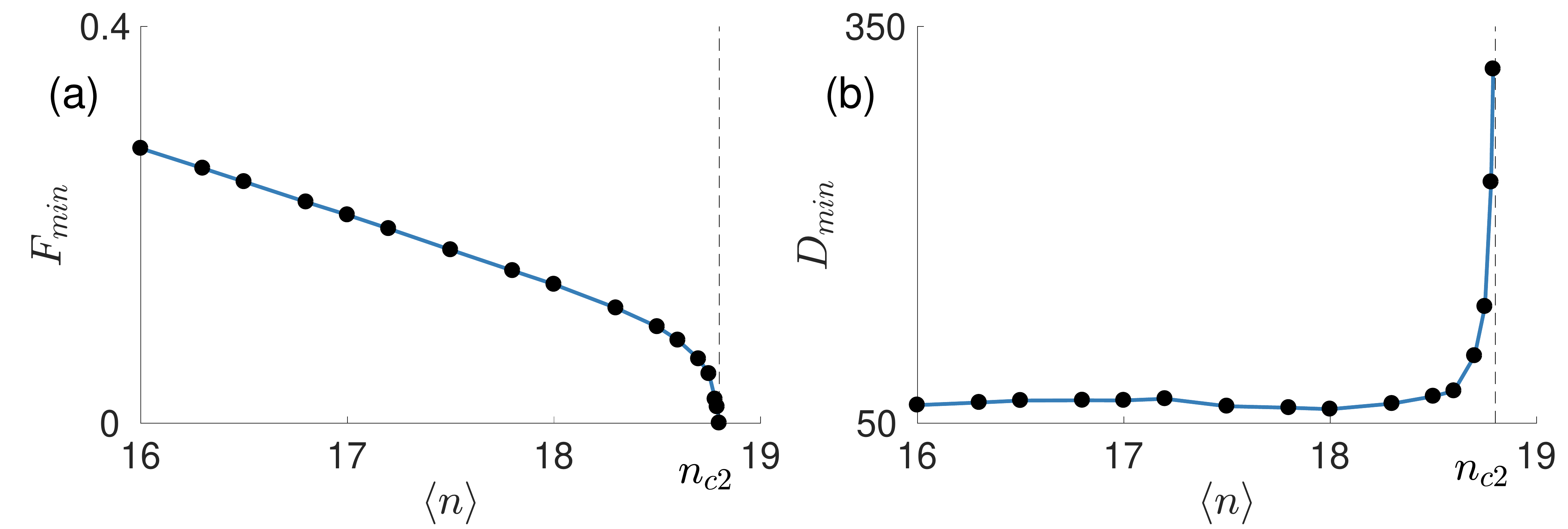}
\caption{Panels (a) and (b) depict the minimum signal amplitude $F_{min}$ to elicit an ILS and the minimum time between two consecutive ILS $D_{min}$ as function of the endogenous stimulation $\langle n \rangle$, respectively.
Parameters in the numerical integration of the rate equations~(\ref{rho-eq}) are the same as in Fig.~\ref{spike_nullclines-fig}.
\label{duration_F_n}}
\end{figure}

If a control parameter such as the endogenous stimulation $\langle n \rangle$ changes slowly from the normal state towards the ictal state through the interictal state, changes in the neuronal dynamics can inform on how close is the system to the critical point of the transition to the ictal state. However, if the variation of the control parameter is too fast to observe its consequences in the dynamics, the transition cannot be anticipated \cite{Silva_2003}. In the SCANNM, since the threshold decreases towards the critical point (Eq.~(\ref{activation-eq})), finite-size effects are more likely to generate ILS, and therefore the rate of ILS is expected to increase. However, as the neuronal network approaches the critical point $n_{c2}$, the stable point R gets closer to the saddle point T. The time $D$ that the system spends in this region of the phase plane increases as
\begin{equation}
D \propto (n_{c2}-\langle n \rangle)^{-1/2}.
\label{D_increase}
\end{equation}
This equation describes a general feature of a SNIC bifurcation \cite{Strogatz_1994}. Therefore, although it becomes easier to elicit ILS from the rest state as the system approaches the critical point, the refractory time also increases, i.e., it takes longer to invoke consecutive ILS. To study the generation of ILS as function of the distance to $n_{c2}$, we stimulate the neuronal network with an excitatory force. In Eqs.~(\ref{rho-eq}) we introduce an additional term $(1-\rho_e)F$ that stimulates the inactive excitatory population $(1-\rho_e)$ with a delta-like field of amplitude $F$ (duration equal to a time step). We found the minimum amplitude $F_{min}$ required to elicit ILS as function of endogenous stimulation, and measured the minimum time between two consecutive ILS $D_{min}$ generated by two consecutive pulses $F_{min}$. Figure~\ref{duration_F_n} shows that $D_{min}$ diverges as the system approaches $n_{c2}$, as Eq.~(\ref{D_increase}) predicts, and $F_{min}$ tends to zero. This is because $F_{min}$ is essentially a measure of the activation threshold, and as such
\begin{equation}
F_{min} \propto \sqrt{n_{c2}-\langle n \rangle}.
\end{equation}
For a sufficiently small fixed stimulation $F$, the neuronal network generates ILS if $F\ge F_{min}(\langle n \rangle)$. The stimulation dependence of $D_{min}$ is a consequence of critical slowing down near $n_{c2}$. (Note that ``critical slowing down'' is a phenomenon by which a system takes longer and longer time to recover from small perturbations as it approaches a critical point of a continuous phase transition \cite{Scheffer_2009}.) At the critical point, $F_{min}\to0$ and $D\to \infty$, meaning that network oscillations emerge with zero frequency, which corresponds to the homoclinic orbit mentioned above. 

\begin{figure}
\includegraphics[width=0.45\textwidth]{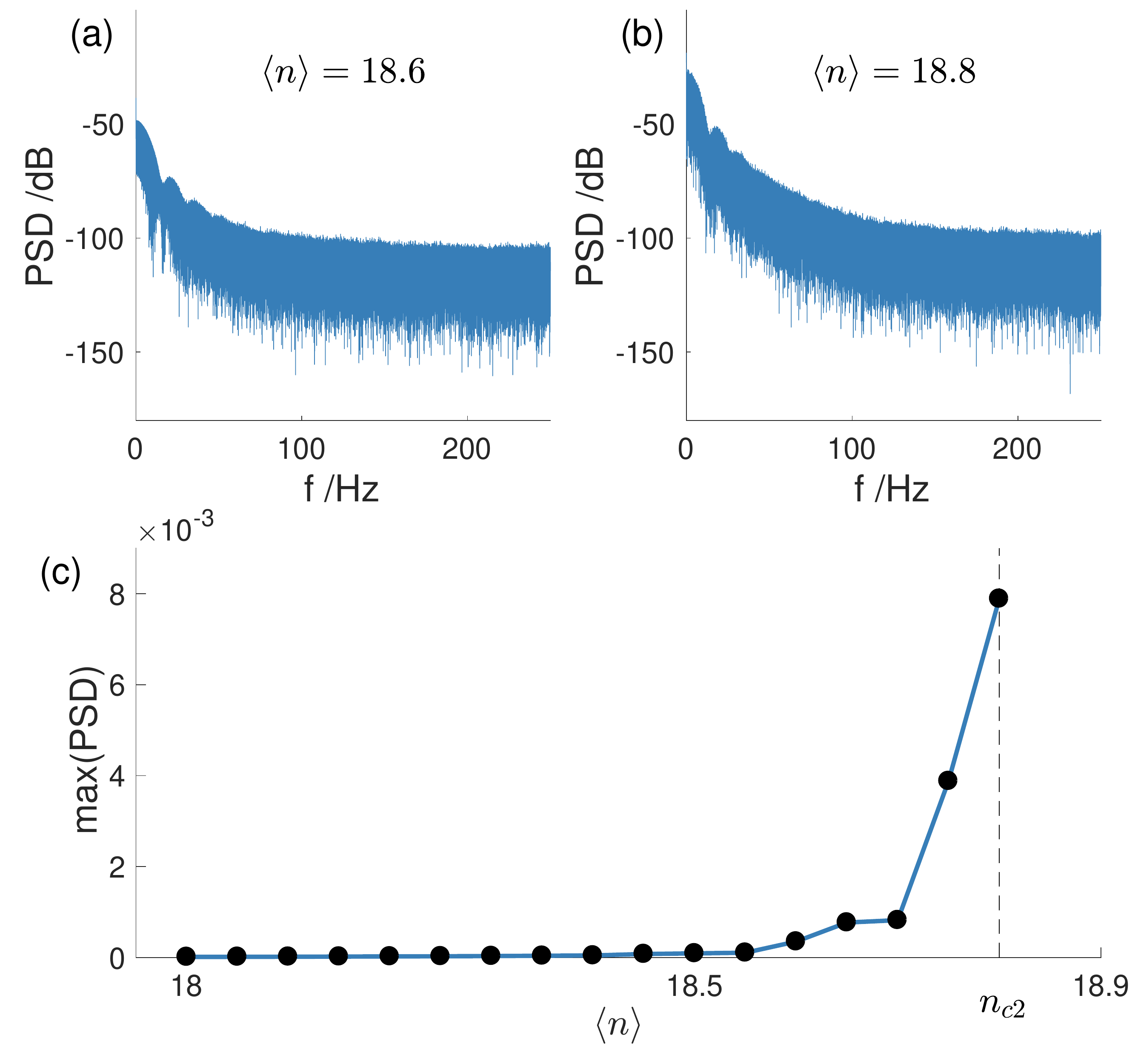}
\caption{Zero-frequency power enhancement as a precursor of the transition to the ictal state. 
Panels (a) and (b) show the power spectral density of activity fluctuations in decibel ($10\log_{10}(\text{PSD})$) at $\langle n \rangle = 18.6$ and $\langle n \rangle = 18.8$, respectively. (c) Maximum of the power spectral density at zero frequency as function of the endogenous stimulation. Parameters in simulations are the same as in Fig.~\ref{spike_nullclines-fig}. 
\label{zero_freq}}
\end{figure}

Besides these dynamical changes involving ILS, the low activity state is also affected by the critical slowing down. This can be quantified by power spectral analysis of low activity fluctuations near the critical point. Using the Wiener-Khintchine theorem, we have demonstrated that the power spectral density (PSD) of activity fluctuations in the low activity state when $\langle n \rangle \to n_{c2}$ has a sharp zero-frequency peak which grows as $S_{max}\propto 1/(n_{c2}-\langle n \rangle)$ \cite{Lee_2014}. This behavior was demonstrated in a metastability region in the vicinity of a first-order phase transition in Ref.~\cite{Lee_2014}, and it also occurs near the second-order phase transition under consideration. Figure~\ref{zero_freq} shows that the zero-frequency peak of the PSD increases as the neuronal network approaches the critical point $n_{c2}$.

\section{Model of recurrent transitions to the ictal state}
\label{recurrent}
\begin{figure}
\includegraphics[width=0.49\textwidth]{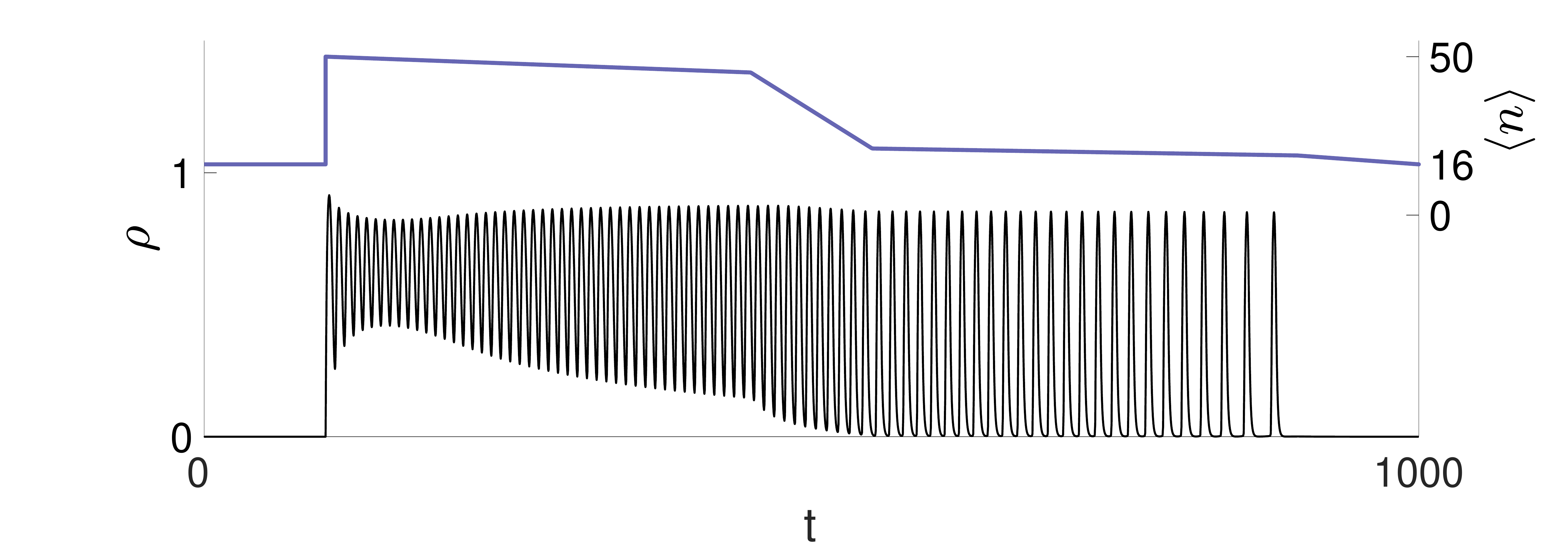}
\caption{Ictal-like pattern evolution driven by endogenous stimulation in the SCANNM. The blue line is time-dependent stimulation $\langle n \rangle$ that varies in the range $[16,50]$ (right y-axis) and drives the neuronal activity $\rho$ (black line) from the low activity state to fast low-amplitude oscillations with a DC shift, which then evolves to high-amplitude, low-frequency oscillations, before returning to the low activity state. Parameters in the numerical integration of the rate equations~(\ref{rho-eq}) are the same as in Fig.~\ref{spike_nullclines-fig}.
\label{seizure_evolution}}
\end{figure}

If we relax the condition of a slowly changing control parameter, the SCANNM is capable of mimicking a typical ictal pattern evolution \cite{Markand_2003,Jirsa_2014}. Figure~\ref{seizure_evolution} shows that an abrupt increase of the endogenous stimulation can bring the neuronal network from the low activity state to the vicinity of the supercritical Hopf bifurcation, which results in a DC shift of the neuronal activity accompanied by fast low-activity oscillations. As then $\langle n \rangle$ slowly decreases back to the 'normal' or the interictal state, the frequency of the sustained network oscillations decreases, whereas its amplitude increases. 

\begin{figure}
\includegraphics[width=0.45\textwidth]{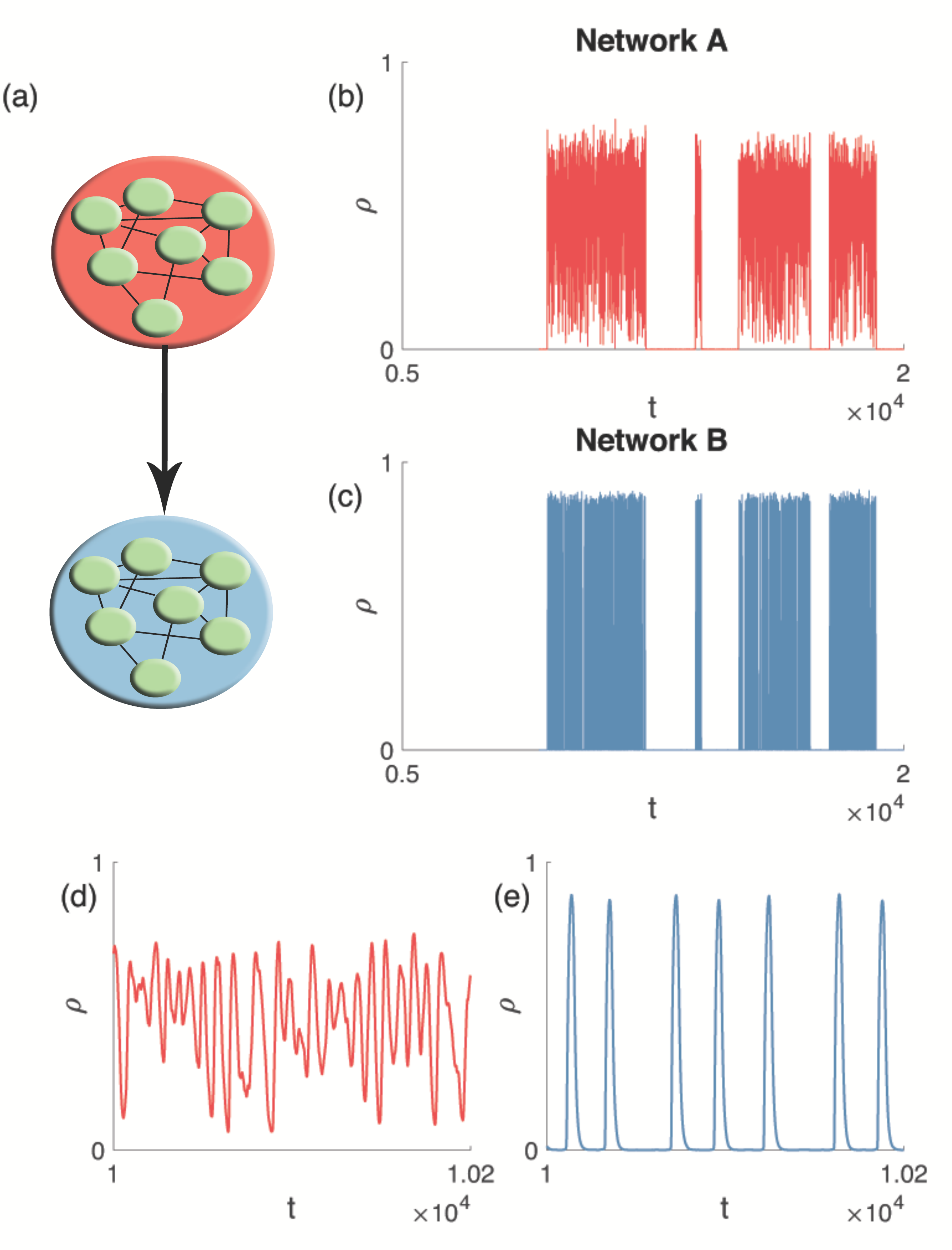}
\caption{Minimal model of recurrent ictal transitions. 
(a) Schematic representation of two networks, where the red network (A) influences the blue (B). 
(b) Neuronal activity in network A. (c) Neuronal activity in network B. The clusters of ictal-like activity in network B are driven by the noisy bursts of network A. 
Panels (d) and (e) are zooms of the activity displayed in panels (b) and (c) respectively. 
Parameters in simulations are the same as in Fig.~\ref{spike_nullclines-fig}, except for those referred in the text.
\label{recurrent_seizures}}
\end{figure}

At fixed parameters, the SCANNM can either be in the normal, interictal, or ictal state. Recurrent transitions between these states can either be achieved by a change in parameters, or due to external stimuli. Another scenario is to consider a network of networks, i.e., several interacting neuronal networks. In general, a network of networks can consist of multiple networks whose inter-network connections can be both directed or undirected, connecting different numbers of excitatory and inhibitory neurons. We consider two interacting neuronal networks as our minimal model of different interacting brain areas. If one of the networks is in the interictal state, then a small additional excitatory input from another network can induce a transition to the ictal state. Recurrent transitions will occur as a consequence of a recurrent input. Such intermittent input can be generated by a network in the bursting state (see Fig.~\ref{diagram-fig}). As previously described in Ref.~\cite{Lee_2014}, the SCANNM produces recurrent irregular bursts of neuronal activity when close to a first-order phase transition. To illustrate the concept, we consider two networks A and B of size $N=10^4$. Network A is in the bursting state [$(\langle n \rangle,\alpha)=(18.7,0.85)$, see the Appendix \ref{appendix_SCANNM} for the meaning of $\alpha$] and network B is in the interictal state [$(\langle n \rangle,\alpha)=(18,0.7)$]. For simplicity, network A sends axonal projections to network B, but network B does not influence network A (see Fig.~\ref{recurrent_seizures}(a)). We define a fraction $g_{AB}=0.3$ of excitatory neurons chosen at random in each network, and we wire them by directed connections from A to B (synaptic efficacies $J_{AB}=3J_e$, see the Appendix \ref{appendix_SCANNM}). In Figs.~\ref{recurrent_seizures}(b) and (c), we show the neuronal activity of the two networks. The recurrent transitions to the ictal state in network B are driven by the bursting activity in network A. 

\section{Discussion}
\label{discussion}
In this paper we proposed a neuronal network model to describe interictal spikes and recurrent ictal activity. ILS are strongly nonlinear collective events that comprise the synchronized activity of a large number of neurons. ILS emerge from a low background activity when either random fluctuations or stimuli force the neuronal activity to overcome a threshold (see Fig.~\ref{spike_nullclines-fig}). This threshold becomes smaller as a control parameter (an endogenous stimulation) moves the neuronal network towards the transition to the ictal state. The transition corresponds to a SNIC bifurcation at which sustained network oscillations emerge with low frequency. This region of oscillations is also bounded at higher endogenous stimulations by a supercritical Hopf bifurcation. Near the Hopf bifurcation, oscillations have low-amplitude and high-frequency. 

It is conceivable that the transition to seizures may either be a consequence of a gradual or an abrupt change in the endogenous stimulation depending on the type of epilepsy \cite{Silva_2003}. In the SCANNM, if we assume a gradual increase of the endogenous stimulation, then we can observe critical phenomena that signal the transition. This transition can be compared to the high amplitude slow (HAS) activity onset pattern observed in some focal epilepsies (see \cite{Wang_2017} and references therein). We showed that under this assumption, the ILS activation threshold scales as the square root of the distance to the SNIC bifurcation, whereas the minimum time between two consecutive spikes diverges at the critical point of the transition to the ictal state (see Fig.~\ref{duration_F_n}). Additionally, the zero-frequency peak of the PSD of low activity fluctuations reaches a maximum at the critical point (see Fig.~\ref{zero_freq}). 

Alternatively, if we instead assume that the control parameter may change abruptly, then the SCANNM is capable of mimicking the typical pattern evolution observed in seizures (called low amplitude fast (LAF) activity onset pattern in \cite{Wang_2017}). Seizures are often preceded by a low-voltage, high-frequency discharge \cite{Wendling_2003}, and the ictal pattern generally displays increasing amplitude and decreasing frequency \cite{Markand_2003}. The SCANNM exhibits such pattern evolution if we assume that at seizure onset the endogenous stimulation abruptly increases, forcing the neuronal network to jump from the interictal-like state to the vicinity of the Hopf bifurcation. Then, as the stimulation gradually decreases towards 'normal' levels, the neuronal activity evolves from low-amplitude, fast oscillations, to high-amplitude, slow oscillations (see Fig.~\ref{seizure_evolution}). 

Finally, we demonstrated the viability of modelling recurrent ictal transitions using two interacting neuronal networks, where the intermittent output of one network drives the other to recurrent seizures. This concept aligns well with other modelling approaches which have also explored the role of interacting populations to generate ictal-like activity \cite{Goodfellow_2011,Goodfellow_2013,Rothkegel_2011,Rothkegel_2014}. Although for simplicity we have considered here the mechanism to seizures as an excitatory drive from another network, we would like to note that this is equivalent to an interruption of inhibition. Such mechanism of triggering seizure-like activity has been observed in genetically engineered mice, where the shut-down of CA2 output leads to hyperexcitability in the recurrent CA3 network \cite{Boehringer_2017}. Note however that the networks involved in the generation of ictal activity may be located in distant regions of the brain. This highlights the importance of studying large scale brain networks, rather than focal brain activity even in the case of focal epilepsies \cite{Richardson_2012}. 

Contrarily to previous computational models of epilepsy \cite{Wendling_2002,Jirsa_2014,Wendling_2016}, the SCANNM was not explicitly designed to describe epileptiform activity \cite{Goltsev_2010,Lee_2014}. Instead, this model demonstrates that interictal and ictal-like activity may be emergent phenomena of an interacting neuronal network. The heterogeneous mean-field equations of the SCANNM were derived from a minimal set of fundamentals, namely neurons behave as stochastic integrate-and-fire neurons, there are two types of neurons (excitatory and inhibitory), and the neurons interact on a complex network \cite{Goltsev_2010,Holstein_2013,Lee_2014}. Our analytical and numerical results were in good agreement with simulations of the model for sufficiently large networks (typically for $N>10^3$ \cite{Goltsev_2010}). We conclude that the SCANNM thus allows to study numerically and analytically neural mass-like equations, and to measure and compare single neuronal activity. It enables to examine how the activity of single neurons can impact on the whole network. For the case in point, ILS are a remarkable example of a collective network phenomenon that can be evoked by the simultaneous activation of a few neurons. On the other hand, once an ILS is excited, the SCANNM predicts that it is very difficult to suppress it. The only way to make the neuronal network return to the low activity state is by applying a strong inhibitory stimulus to a considerable macroscopic part of the network. 

A recent study has reported apparently self-contradictory evidence on the role of pre-ictal spikes for the prediction of seizures \cite{Karoly_2016}, showing that different seizures could be preceded by an increase or decrease of the pre-ictal spike rate. The SCANNM provides a possible explanation for this observation. In the model, as a transition to the ictal state is approached, two competing mechanisms can influence the spike rate. On one hand, the activation threshold of ILS decreases which leads to a higher spike rate. On the other hand, critical slowing down hinders the consecutive emergence of spikes (see Fig.~\ref{duration_F_n}). It is then conceivable that the prevailing mechanism may vary from seizure to seizure, and as a result the spike rate can increase or decrease before a seizure.

There is also conflicting evidence as to which it may or may not be possible to predict seizures based on critical phenomena \cite{Milanowski_2016} or using other data analysis \cite{Mormann_2007}. Nevertheless, based on the SCANNM we can propose two measures to forecast seizures. First, Fig.~\ref{duration_F_n}(a) indicates that the required external stimulation to evoke ILS becomes smaller as the neuronal network approaches the transition to the ictal state. In the case of photosensitive epilepsy, the stimulation can be intermittent photic stimulation. In other epilepsies it may be necessary to use implanted electrodes to electrically stimulate the brain, like it was proposed by Silva \textit{et al}. \cite{Silva_2011}. Thus, for a given patient, and after a sufficient number of trials, it may be possible to correlate the minimum required stimulation to elicit ILS with the timing of impending seizures. However, such method may be infeasible due to the risk of inducing seizures due to this probing stimulation \cite{Kalitzin_2010}. The second measure does not require stimulation, instead it uses the analysis of ongoing EEG recordings. Figure~\ref{zero_freq} shows that the power of low frequencies should increase in EEG recordings when approaching a transition to a seizure. In fact, a gradual increase in power of low frequencies has been observed as a precursor of spike-wave discharges in absence epilepsy both in humans and rat models \cite{Inouye_1990,Gupta_2011,Sitnikova_2009,Luijtelaar_2011}. We acknowledge, however, that even if this process takes place, it may often be unobservable because if the transition occurs in a faster time scale than the scale of the low frequencies, then it is not possible to find the gradual power increase.

\section{Conclusion}
In this paper we proposed a neuronal network model (the SCANNM) to describe interictal and ictal activity, as well as ictal-like pattern evolution, and spontaneous recurrent transitions to seizures. Additionally, we found a set of precursors that signal the transition to the ictal state. The neuronal activity state was dependent of an endogenous stimulation which we used as the control parameter. The interictal state was characterized in the model by low fluctuating activity from which interictal-like spikes could sporadically emerge. We demonstrated that the required stimulation to elicit interictal-like spikes tends to zero as the neuronal network approaches the critical point of a saddle-node bifurcation. Furthermore, the transition is signaled by an increase of the zero-frequency peak of the power spectrum of low activity fluctuations when the endogenous stimulation varies slowly. On the other hand, for abrupt changes in the control parameter, we showed that the model can mimic a typical ictal pattern evolution: as onset with low-voltage, high-frequency discharges, followed by increasing amplitude, decreasing frequency oscillations. Finally, we showed that the model could also reproduce recurrent transitions to the ictal state at fixed parameters, as the result of the interaction between two neuronal networks.

\section{Acknowledgements}
This work was partially supported by FET IP Project MULTIPLEX 317532. A.V.G. is grateful to LA I3N for Grant No. PEST UID/CTM/50025/2013. M.A.L. acknowledges the financial support of the Medical Research Council (MRC) via grant MR/K013998/01. K.E.L. was supported by the Department of Anesthesiology at the University of Michigan (Ann Arbor) and National Institutes of Health (Bethesda, MD, USA) grant RO1 GM098578.

\appendix
\section{SCANNM}
\label{appendix_SCANNM}
Here we describe the SCANNM \cite{Goltsev_2010,Lee_2014,Lopes_2014}. 
 
\subsection{Network structure and stochastic dynamics}
\label{structure}
The neuronal network is composed of $N$ stochastic neurons, $g_eN$ excitatory and $g_iN$ inhibitory ($g_e+g_i=1$). We consider that the network has the structure of the Erd\H{o}s-R\'{e}nyi network. This is a random network with small world properties, namely small mean shortest path length like real neuronal networks in the brain \cite{Sporns_2004}. The neurons are connected by directed edges which represent synapses that allow active neurons to send spikes to their postsynaptic neighbors. In addition, neurons also receive random spikes from endogenous stimulation that represent spontaneous releases of neurotransmitters in synapses and random spikes arriving from other areas of the brain (this stimulation has properties of shot noise \cite{Lee_2014}). 

The dynamics of the stochastic neurons is determined by the following rules. If during an integration time window $\tau$ the total input $V_j(t)$ to an inactive neuron becomes larger than a threshold $\Omega$, then with probability $\tau \mu_{a}$ the neuron becomes active and fires a spike train at a constant frequency $\nu$ (the index $a=e$ if the neuron is excitatory and $a=i$ if it is inhibitory). If the total input $V_j(t)$ to an active excitatory (inhibitory) neuron becomes smaller than $\Omega$, then the neuron stops to fire with probability $\tau \mu_{a}$. In this model, the rates $\mu_{e}$ and $\mu_{i}$ are the reciprocal first-spike latencies  of excitatory and inhibitory neurons, respectively. We define a parameter $\alpha$ as the ratio of the first-spike latency of excitatory neurons to the first-spike latency of inhibitory neurons, $\alpha \equiv \mu_{i}/\mu_{e}$. If $\alpha < 1$, then excitatory neurons respond faster to stimuli than inhibitory neurons. This neuronal stochastic behavior is meant to account for intrinsic noise within neurons \cite{Mainen_1995}, namely, ion channel stochasticity \cite{Schneidman_1998}.

\subsection{Rate equations}
\label{rate_equations}
The fractions $\rho_e(t)$ and $\rho_i(t)$ of active excitatory and inhibitory neurons, respectively, characterize the neuronal activity at time $t$. They are determined by the rate equations (\ref{rho-eq}), in which $\Psi_a(\rho_e,\rho_i,\langle n \rangle)$ is the probability that, at time $t$, the total input to a randomly chosen excitatory ($a=e$) or inhibitory ($a=i$) neuron is at least the threshold $\Omega$ at a given endogenous stimulation $\langle n \rangle$. The functions $\Psi_a(\rho_e,\rho_i,\langle n \rangle)$ are determined by the network structure, the distribution function of endogenous stimulation (we consider the Gaussian distribution), and the frequency-current relationship of single neurons (a step function in this model \cite{Goltsev_2010}). Note that the probability $\Psi_{a}(\rho_e,\rho_i,\langle n \rangle)$ is the same for both excitatory and inhibitory neurons because, in the network under consideration, excitatory and inhibitory neurons occupy topologically equivalent positions. Thus, $\Psi_e(\rho_e,\rho_i,\langle n \rangle)=\Psi_i(\rho_e,\rho_i,\langle n \rangle)\equiv \Psi(\rho_e,\rho_i,\langle n \rangle)$, where
\begin{eqnarray}
&\Psi(\rho_e,\rho_i,\langle n \rangle)= \sum_{n,k,l\geq 0} \Theta( nJ_{n}{+}kJ_e{+}lJ_i{-}\Omega)G(n,\langle n \rangle) \times \nonumber \\
& P_k(g_e\rho_e\tilde{c})P_l(g_i\rho_i\tilde{c}).
\label{psi-eq}
\end{eqnarray}
Here, $\tilde{c}=c \nu \tau$ and $c$ is the mean degree. $\Theta(x)$ is the Heaviside step function. $P_{q}(c)$ is the Poisson distribution function,
\begin{equation}
P_{q}(c)=c^{q}e^{-c}/q!,
\end{equation}
and $G(n,\langle n \rangle)$ is the Gaussian distribution function,
\begin{equation}
G(n,\langle n \rangle)=G_0 e^{-(n-\langle n \rangle)^2/2\sigma^2}.
\label{g-noise}
\end{equation}
$G(n,\langle n \rangle)$ defines the probability that a neuron receives $n$ spikes from endogenous stimulation per integration time $\tau$. $\langle n\rangle$ is the mean number of spikes, $\sigma^2$ is the variance, and $G_0$ is the normalization constant, $\sum_{n=0}^{\infty}G(n,\langle n \rangle)=1$. We use $\langle n\rangle$ as the control parameter characterizing the endogenous stimulation. Note that Eqs. (\ref{rho-eq}) and (\ref{psi-eq}) are asymptotically exact in the thermodynamic limit, $N \rightarrow \infty$ \cite{Goltsev_2010,Lee_2014}.

In numerical simulations, we use the algorithm explained in \cite{Lee_2014}. We used the following model parameters (except when mentioned otherwise): $N=10^4$, $c=10^3$, $\Omega=30$, $\tau \nu = 1$, $\mu_{e} \tau =0.1$, $\alpha=0.7$, and $g_i=0.25$. Throughout this paper we used $1/\mu_{e}\equiv 1$ as time unit and $J_e\equiv 1$ as input unit. Following \cite{Amit_1997}, we chose $J_{i}=-3J_{e}$. We also used $J_{n}=J_{e}$ and $\sigma^2 =10$ for the amplitude and variance of the endogenous stimulation. These parameter choices have been discussed in \cite{Goltsev_2010,Lee_2014,Lopes_2014}.

\section{The activation threshold of ILS}
\label{app_activation_th}
In this appendix we show how the activation threshold $A_{th}$ of ILS depends on the endogenous stimulation $\langle n \rangle$ near the SNIC bifurcation. Since we consider $\Psi_e=\Psi_i\equiv\Psi$, this implies that there is only one steady state equation,
\begin{equation}
\rho=\Psi(\rho,\langle n \rangle).
\label{steady-eq}
\end{equation}
In this case, the SNIC bifurcation occurs when 
\begin{equation}
\frac{\partial \Psi}{\partial \rho}(\rho,\langle n \rangle)=1, 
\label{condition-eq}
\end{equation}
which determines the critical point $n_{c2}$. In the region of ILS, close to the bifurcation, at $0<n_{c2}-\langle n \rangle \ll n_{c2}$, we can study low activity fluctuations $\delta\rho(\langle n \rangle)=\rho(\langle n \rangle)-\rho(n_{c2})$ near the SNIC bifurcation. To find how the activity fluctuations depend on the stimulation $\langle n \rangle$, we apply the Taylor expansion to $\Psi(\rho,\langle n \rangle)$ in Eq.~(\ref{steady-eq}) over $\delta\rho(\langle n \rangle)$ and $\delta n=\langle n \rangle-n_{c2}$ up to the second order in $\delta\rho(\langle n \rangle)$,
\begin{equation}
\delta\rho(\langle n \rangle)\approx \frac{\partial \Psi}{\partial \langle n \rangle}\delta n+\frac{\partial \Psi}{\partial \rho}+\frac{\partial^2 \Psi}{\partial \rho^2}(\delta \rho)^2,
\end{equation}
where the derivatives of $\Psi$ are taken at $n_{c2}$. Using Eq.~(\ref{condition-eq}), we find
\begin{equation}
\rho(\langle n \rangle)-\rho(n_{c2}) \approx - C \sqrt{n_{c2}-\langle n \rangle}, 
\end{equation}
where
\begin{equation}
C=\sqrt{-2\frac{\partial\Psi}{\partial \langle n \rangle}\Big(\frac{\partial^2\Psi}{\partial \rho^2}\Big)^{-1}}. 
\end{equation}
Consequently, near the critical point $n_{c2}$, the activation threshold $A_{th}(\langle n \rangle)$ also follows
\begin{equation}
A_{th}(\langle n \rangle) \propto \sqrt{n_{c2}-\langle n \rangle}.
\end{equation}

\bibliography{Lopes_bib}
\end{document}